\documentclass[prb, twocolumn,amsmath,amssymb]{revtex4}
\usepackage{graphicx, color}
\usepackage{dcolumn}
\usepackage{bm}
\usepackage{epstopdf}
\usepackage{color}
\usepackage{subfigure}
\begin{document}
\title{Orbital Limit and Gaussian Fluctuation Effects in Flat Band Superconductors with PseudoMagnetic Fields} 

\author{Xiao-Hui Li}
\author{Yao Lu}
\thanks{yluae@connect.ust.hk}
\affiliation{Department of Physics, Hong Kong University of Science and Technology, Clear Water Bay, Hong Kong, China}
\date{\today}
\begin{abstract}
In this work, we study a molecular graphene model on the top of a superconductor in the presence of pseudomagnetic fields induced by coplanar strain fields. With the pseudomagnetic fields and the attractive interaction induced from the substrate, a flat band superconductor can be achieved according to mean field analysis on the effective Hamiltonian.  Based on a semiclassical approximation, we first show that the orbital limit is  hugely enhanced by the pseudomagnetic fields. The physical reason is that the orbital angular momenta locking at $K$ and $K'$ valleys due to the pseudomagnetic fields suppress the orbital magnetization from external magnetic fields. Considering the vanishing band width in this system, we then study the effects of Gaussian fluctuations in both Hartree and pairing channels. We show that in the dilute limit, the phase transition is dominated by collective modes with critical temperatures much lower than the mean field results. At half filling, our method gives no corrections to the mean field critical temperatures.
\end{abstract}

\maketitle
\section{Introduction}

A flat band is a dispersionless Bloch band with vanishing band width, whose ground state is insulating in the absence of disorder and interaction at any filling [\onlinecite{Resta}]. It has been one of the popular topics in condensed matter physics, as it is closely related to some highly novel phenomena. One prime example is the fractional quantum Hall effect [\onlinecite{Gossard}], in which the Coulomb interaction together with Landau levels (LL) leads to the fractional Hall conductance [\onlinecite{Jain}]. Another interesting example is the magnetism arising from time reversal symmetry (TRS) preserved flat band systems, which has been widely studied in Lieb lattices [\onlinecite{Mielke,Lieb,Tasaki,Tasaki2,Tasaki3}]. Recently, the combination of  TRS preserved flat bands and superconductivity has attracted much interest.  One of the reasons is that it has good perspective for achievement of high temperature superconductivity. According to the Bardeen-Cooper-Schrieffer (BCS) theory [\onlinecite{Schrieffer,Schrieffer2}], with fixed interaction strength, the flat dispersion corresponds to the maximum critical temperature due to the diverging density of states near the Fermi surface [\onlinecite{Yamashita}]. In addition to the enhancement of superconductivity, a flat band superconductor also has its own refreshing properties beyond the BCS framework. For instance, the superfluid weight responsible for zero resistance and diamagnetic effect is proportional to the integral of quantum metric over the Brillouin Zone despite of zero Fermi velocity [\onlinecite{Torma,Torma2,Huber,Torma3}].

Motivated by these fascinating effects, plenty of theoretical models to realize TRS preserved
flat band have been proposed, e.g. Lieb lattices, surfaces of topological nodal insulators [\onlinecite{Fu2,Volovik,Kopnin,Heikkila1,Barlas,Scalettar,Volovik2,Heikkila2,Fu}]. Among these proposals, a very interesting one is to realize LLs in graphene systems or on the surfaces of topological crystalline insulators in the presence of pseudomagnetic fields, in which superconductivity was claimed to arise simultaneously [\onlinecite{Heikkila2,Fu}].  Experimentally, LLs have been observed in graphene systems with strain induced pseudomagnetic field [\onlinecite{Crommie}]. Apart from applying strain directly on graphene, another feasible way to realize pseudomagnetic field is to manipulate molecular lattices [\onlinecite{Manoharan}]. The signatures of flat bands have been observed in a recent experiment, in which deposing CO molecules on the top of copper surface results in Dirac dispersion and deforming the molecular lattice induces the pseudomagnetic fields [\onlinecite{Manoharan}]. By replacing the substrates with superconducting thin films, superconductivity can be introduced into the flat band systems. Consequently, flat band superconductors with high critical temperatures can be realized in molecular graphene with strain fields.

In this work, we study the properties of the flat band superconductors realized by molecular graphene with superconductivity and pseudomagnetic fields. Starting from an effective two dimensional Dirac Hamiltonian, we find the orbital limit roughly proportional to the strength of pseudomagnetic fields. With large pseudomagnetic fields, the orbital limit can be therefore hugely increased. This effect is analogue to the Ising protection of superconductivity from external Zeeman field in an Ising superconductor [\onlinecite{Ye,Mak,Iwasa}]. Noticing the ratio between the interaction strength and band width is extremely large, it is expected that fluctuation effects play a  crucial role at finite temperatures in the dilute limit [\onlinecite{Zwerger,Engelbrecht,Melo,Castro,Randeria,Devreese,Ohashi,Toigo,Liu,Salasnich}]. To quantify the fluctuation effect, we go beyond mean field analysis by applying the extended Nozieres and Schmitt-Rink (NSR) method to calculate the order parameters. We show that at zero temperature, the mean field results match the extended NSR results as expected. However at finite temperature, fluctuation effects become more important. In the dilute limit, the critical temperature, at which the local order parameter vanishes, is hugely suppressed by the fluctuation effects. At the critical point, in the number equation, the main contribution to the particle number comes from fluctuation effect, in which sense it is very similar to Bose-Einstein condensate (BEC) superconductors. However, at half filling, the critical temperatures are almost the same for both cases: with and without fluctuation effects.

\section{Model Hamiltonian}

%

\begin{figure}[h!]
\centering
\subfigure{\includegraphics[width = 0.95\columnwidth]{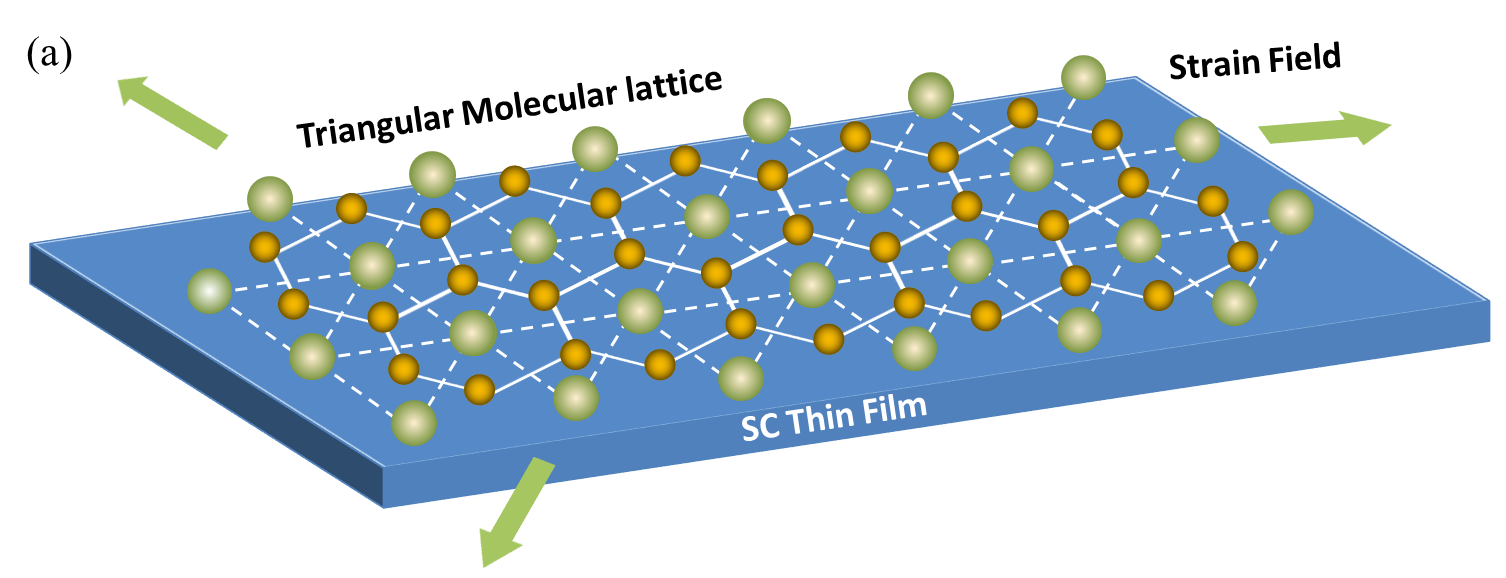}}
\subfigure{\includegraphics[width=0.6\columnwidth]{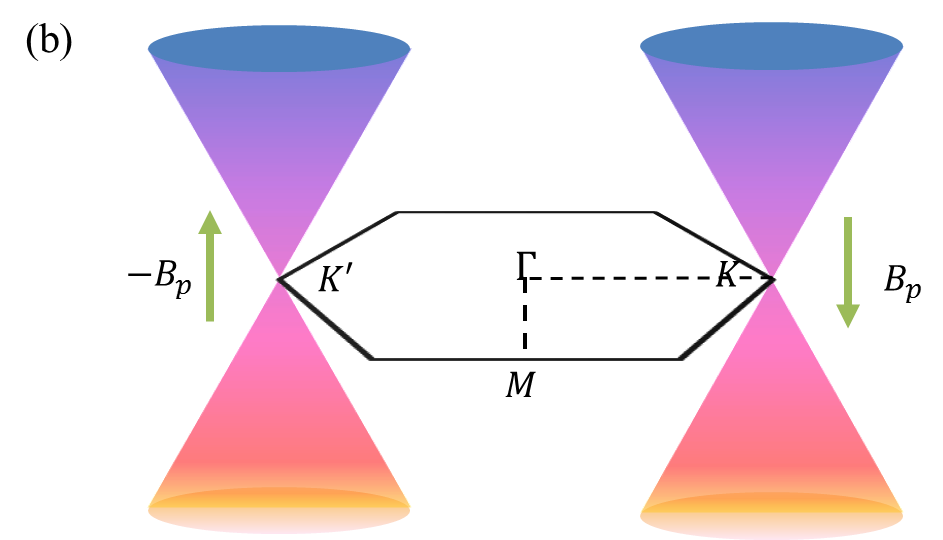}}
\caption{(a)Simplified sketch of the model discussed in the maintext.  Molecules (bigger green spheres) are placed as triangular lattice on the top of a superconducting thin film, consequently the electrons (smaller yellow spheres) are squeezed into the middle region and form honeycomb lattice. The pseudomagnetic field can be induced by  three coplanar 120$^\circ$ strain fields. (b) The Brillouin zone of the formed honeycomb lattice with the high symmetric points ($K$, $\Gamma$, $M$, $K'$) pointed out. The normal state low energy physics lies at $K$ and $K'$ points, described by a Dirac Hamiltonian. With the strain fields, electrons at $K$ and $K'$ valleys experience opposite pseudomagnetic fields $B_p$ and $-B_p$ along $z$ direction.}\label{Fig:Demo}
\end{figure}

Inspired by the recent experiment realizing LLs in molecular graphene system [\onlinecite{Manoharan}], we consider a similar system by assuming the substrate to be superconductor thin film as shown in Fig.\ref{Fig:Demo}(a). As the electrons are squeezed by the molecules and form a honeycomb lattice, Dirac cones appear at around $K$ and $K'$ points (shown in Fig.\ref{Fig:Demo}(b)). In the presence of three coplanar 120$^\circ$ strain fields on the molecular lattice, electrons at $K$ and $K'$ valleys experience opposite out of plane pseudomagnetic fields $B_p$ and $-B_p$ restricted by TRS [\onlinecite{Guinea1,Guinea2}]. Since the substrate is a superconducting thin film, the attractive interaction also plays an important role in the whole system. Hence, the effective Hamiltonian describing the system can be written as  

\begin{equation}
H=H_0+H_{int}=\int d^2\bm{r}\psi^{\dag}(\bm{r})h_{0}\psi({\bf r}) + H_{\mathrm{int}}
\end{equation}
with
\begin{eqnarray}
h_{0}&=&v_F(-i\nabla_x)\sigma_z\gamma_x+v_F(-i\nabla_y)\gamma_y-eA_{px}\gamma_x \nonumber \\
&-&eA_{py}\sigma_z\gamma_y-\mu
\end{eqnarray}
and
\begin{eqnarray}
H_{\text{int}}=-V\sum_{\sigma,\gamma}\int d^2\bm{r}\hat{\Omega}_{\sigma\gamma\tau}(\bm{r})
\end{eqnarray}
where $\hat{\Omega}_{\sigma\gamma\tau}(\bm{r})=c_{\uparrow\sigma\gamma,\tau}^{\dag}(\bm{r})c_{\downarrow\bar{\sigma}\gamma,\tau}^{\dag}(\bm{r})c_{\downarrow\bar{\sigma}\gamma,\tau}(\bm{r})c_{\uparrow\sigma\gamma,\tau}(\bm{r})$. Here 

\begin{eqnarray}
\psi=&&(c_{\uparrow,K,A},c_{\uparrow,K,B,},c_{,\uparrow,K',A,},c_{,\uparrow,K',B}, \nonumber \\
&&c_{\downarrow,K,A},c_{\downarrow,K,B,},c_{\downarrow,K',A,},c_{\downarrow,K',B,})^{\text{T}}
\end{eqnarray}
with $c_{s,K/K',A/B}$ being the annihilation operators of electrons at $K/K'$ valley with sublattice index $A/B$ and spin index $s$. $\sigma$, $\gamma$ and $s$ are Pauli matrices acting on valley, sublattice and spin space respectively. $\bar{\sigma}$ denotes the valley opposite to $\sigma$ valley. $v_F$ is the Fermi velocity, $\mu$ is the chemical potential away from half filling. $e$ is the charge of an electron. $A_{px}$ and $A_{py}$ are $x$ and $y$ components of pseudomagnetic field vector potential. The pseudomagnetic field couples to electrons like magnetic field but with opposite signs for the two valleys. It preserves TRS and thus is compatible with superconductivity. The interaction strength $-V$ is negative indicating attractive interaction. For convenience, we choose the Landau gauge for the pseudomagnetic field $A_{p,x}=yB_{p},A_{p,y}=0$. By solving the eigenvalue equation, it is easy to obtain the quasi-particle excitation energy $\epsilon_{n,k_x}=v_F\sqrt{2neB_p}$, where $n$ is the LL index, $k_x$ is the momentum of electrons along $x$ direction. Here we assume the LLs are filled up to the $0th$ one.

Within the BCS framework, the interaction Hamiltonian can be decoupled as 

\begin{eqnarray}
H_{\textbf{int}}\approx H_{\Delta}=\sum_{\sigma,\gamma}\int d^2\bm{r}\left(\Delta_{\gamma}c_{\uparrow，\sigma,\gamma}c_{\downarrow，\bar{\sigma},\gamma}+h.c.\right)
\end{eqnarray}
where the order parameter $\Delta_{A/B}=V\sum_{\sigma}\langle c_{\uparrow,\sigma,A/B}(\bm{r})c_{\downarrow,\bar{\sigma},A/B}(\bm{r})\rangle$ should be determined self consistently. And the Hartree contribution from the interaction can be absorbed into the effective chemical potential $\tilde{\mu}$. Therefore, one obtains the mean field Bogoliubov–de Gennes (BdG) Hamiltonian $H_{BdG}=\int d^2\bm{r}\Psi(\bm{r})^{\dag}h_{BdG}\Psi(\bm{r})$, with

\begin{equation}
h_{BdG}=\begin{pmatrix}
h_0   &&   \bm{\Delta}  \\
\bm{\Delta}^{\dag}  && -\Theta h_0\Theta^{-1}
\end{pmatrix},
\end{equation}
where the basis $\Psi(\bm{r})=[\psi(\bm{r}),\psi(\bm{r})^{\dag}]^{\text{T}}$ is the 16 component spinor in Nambu space. $\Theta$ is the time reversal operator. The off-diagonal term $\bm{\Delta}$ is the pairing matrix that describes the formation of Cooper pairs. The mean field order parameter and critical temperature have been calculated in Ref.[\onlinecite{Barlas}] on the uniform pairing condition. In the rest of this article, we always assume the LLs are filled up to the 0$th$ one and the LL spacing is much larger than the critical temperature, such that we can focus on the 0$th$ LL and ignore the higher ones. In this limit, the order parameter at zero temperature can be calculated as $\Delta_A(T=0)=-VN_{\phi}\sqrt{1-4\nu^2}$ ， where $N_{\phi}$ is the LL degeneracy per unit cell and $\nu$ is the filling factor, which is set to be 0 at half filling.

\section{Valley Angular Momentum Locking and Orbital Limit}
In conventional $s$ wave superconductors, a Cooper pair is formed with two electrons carrying opposite spins and angular momenta. When applying a $real$ magnetic field $B_r$ in $z$ direction, the Cooper pairs can be broken by both Zeeman and orbital effects, which pin the spins and angular momenta of the electrons into the same direction, respectively. However, in the system we consider, due to the effect of pseudomagnetic field, electrons at $K$ and $K'$ valleys experience different effective magnetic fields $B_r+B_p$ and $B_r-B_p$. Intuitively thinking, to break the Cooper pairs through orbital effect one needs a $real$ magnetic field $B_r$ comparable with $B_p$ such that $B_r+B_p$ and $B_r-B_p$ have the same sign. To explicitly analyze the orbital limit, we start with the following linearized gap equation [\onlinecite{Werthamer}]

\begin{widetext}

\begin{equation}
\Delta(\bm{r}_2)=-\frac{1}{4}TV\sum_m\int d^2\bm{r}_1\text{Tr}\left[G_0(i\omega_m,\bm{r}_2,\bm{r}_1)\tau_ys_yG_0(-i\omega_m,\bm{r}_1,\bm{r}_2)\tau_ys_y\right]\Delta(\bm{r}_1)
\end{equation},

\end{widetext}
where $\mathbb{\tau}_is$ are the Pauli matrices defined in particle-hole space. And $G_0$ is the Gorkov Green's function with vanishing order parameter, which reads 
\begin{equation}
G_0^{-1}=\partial_{\tau}+h_{BdG}(\Delta=0)
\end{equation}

With semiclassical approximation, we can write the gap equation in the presence of external electromagnetic field as 

\begin{widetext}

\begin{equation}\label{Eq:SemiGap}
\Delta(\bm{r}_2)=-\frac{1}{4}TV\sum_m\int d^2\bm{r}_1\text{Tr}\left[G_0(i\omega_m,\bm{r}_2,\bm{r}_1)\tau_y\sigma_yG_0(-i\omega_m,\bm{r}_1,\bm{r}_2)\tau_y\sigma_y\right]e^{(\bm{r}_2-\bm{r}_1)\cdot(\bm{\nabla}-2ie\bm{A}_{r})}\Delta(\bm{r}_2)
\end{equation}

\end{widetext}
Through a detailed derivation (shown in Appendix A), we obtain the gap equation as the following, 

\begin{eqnarray}\label{Eq:Gap2}
\Delta(\bm{r})=&&V\int d\rho \quad \frac{\rho N_{\phi}}{\tilde{\mu}}\tanh(\tilde{\mu}/2T)\nonumber \\
&& e^{-\frac{1}{2}\rho^2eB_p}e^{-\frac{1}{2}\rho^2eB_c}\Delta(\bm{r})
\end{eqnarray}
where $\tilde{\mu}$ is the effective chemical potential and $B_c$ is the orbital limit. Working out the integral,  the orbital limit yields

\begin{equation}\label{Eq:Bc}
B_c=B_p\left(\frac{V N_{\phi}\tanh(\tilde{\mu}/2T)}{\tilde{\mu}}-1\right)
\end{equation}
Since $\Delta$ is close to zero near phase transition point, the effective chemical potential $\tilde{\mu}$, filling factor $v$ and temperature $T$ have the following relation

\begin{equation}\label{Eq:v}
\nu=\frac{1}{e^{-\tilde{\mu}/T}+1}-\frac{1}{2}
\end{equation}
Combining Eq. (\ref{Eq:Bc}) and (\ref{Eq:v}), we have

\begin{equation}\label{Eq:Bc2}
B_c=B_p\left(\frac{2VN_{\phi}\nu}{T\log(\frac{v+1/2}{1/2-\nu})}-1\right)=B_p\left(\frac{T_c(B_r=0)}{T}-1\right)
\end{equation}

The orbital limit is roughly proportional to the strength of pseudomagnetic field. And the coefficient is solely determined by the ratio between the critical temperature at zero field ($T_c(B_r=0)$) and the temperature $T$. Experimentally, the pseudomagnetic field can be as large as $60T$ [\onlinecite{Manoharan}]. When the temperature is half of the critical temperature $T=\frac{1}{2}T_c(B_r=0$), the orbital limit calculated from Eq. (\ref{Eq:Bc2}) is the same as the pseudomagnetic field $B_c=B_p$. However, this is not the actual case. When $B_r$ approaches $B_p$, the $K$ and $K'$ valleys experience effective magnetic field $2B_p$ and $0$ respectively. $K'$ valley will not form LLs. In this case, we cannot focus on 0$th$ LL and drop the higher ones. Therefore, the conclusion we can draw from our calculation is that, when temperature is higher than $T_c/2$, Eq. (\ref{Eq:Bc2}) gives a good approximation of orbital limit. On the other hand, if the temperature is lower than $T_c/2$, Eq. (\ref{Eq:Bc2}) is no longer valid. But we can still claim that if $T<T_c/2$, the orbital limit has a rough lower bound $B_c\geq B_p$.

\section{Fluctuation Effect at Finite Temperature}

So far, all the calculations are based on mean field theory, which is valid in the conventional BCS superconductors. In flat band systems, the ratio between interaction strength and band width is in the BEC limit. To properly describe this system, we need to take into account the quantum fluctuations in both density and pairing channels. However, it is known that the standard Hubbard-Stratonovich (HS) transformation cannot deal with both kinds of fluctuations simultaneously [\onlinecite{Orland}]. Here we apply the generalized Hubbard-Stratonovich (GHS) transformation approach developed by Kerman [\onlinecite{Kerman1,Kerman2,Kerman3}].  We start with the partition function 

\begin{equation}
Z=\text{Tr}e^{-\beta H}=\lim_{\epsilon\rightarrow 0}\text{Tr}\text{T}_{\tau}\prod_{\tau=1}^{N_{\tau}}\left[1-\epsilon H_{0,\tau}-\epsilon H_{int,\tau}\right]
\end{equation}
where T$_{\tau}$ is the imaginary time ordering operator, $\tau$ is the imaginary time index and $\epsilon=\beta/N_{\tau}$ is the small imaginary time length for each time slice. Introducing auxiliary fields by inserting a fat identity, we can decouple the interaction in both Hartree and Bogoliubov channels following Ref.[\onlinecite{Macdonald,Kerman3}]

\begin{equation}
Z=\lim_{\epsilon\rightarrow 0}\int\prod_{\tau=1}^{N_{\tau}} D[\phi,\Delta,\bar{\Delta}]\exp(-S[\phi,\Delta,\bar{\Delta}])
\end{equation}

\begin{eqnarray}
S[\phi,\Delta,\bar{\Delta}]&=&\epsilon\sum_{\tau=1}^{N_{\tau}}\int d^2\bm{r}\left[\frac{\phi_{\tau}(\bm{r})\phi_{\tau}(\bm{r})}{2U}+\frac{\bar{\Delta}_{\tau}(\bm{r})\Delta_{\tau}(\bm{r})}{U} \right] \nonumber \\
&&-\log \text{T}_{\tau}\text{Tr}\prod_{\tau=1}^{N_{\tau}} \left[1-\epsilon\tilde{H}_{M,\tau}-\epsilon^2 \tilde{H}_{int,\tau}\right]
\end{eqnarray}
with

\begin{eqnarray}
\tilde{H}_{M,\tau}&=&\epsilon H_{0,\tau}-\epsilon\tilde{H}_{\phi,\tau}-\epsilon \tilde{H}_{\Delta,\tau} \nonumber \\
\tilde{H}_{\phi,\tau}&=&\sum_{s\sigma\gamma}\int d^2\bm{r}\frac{1}{\sqrt{2}}\phi_{\tau}(\bm{r})c_{s\sigma\gamma,\tau}^{\dag}(\bm{r})c_{s\sigma\gamma,\tau}(\bm{r}) \nonumber \\
\tilde{H}_{\Delta,\tau}&=&\sum_{\sigma\gamma}\int d^2\bm{r}\Delta_{\tau}(\bm{r})c_{\uparrow\sigma\gamma,\tau}^{\dag}(\bm{r})c_{\downarrow\bar{\sigma}\gamma,\tau}^{\dag}(\bm{r})+h.c. \nonumber \\
\tilde{H}_{int,\tau}&=&\sum_{\sigma\gamma}\int d^2\bm{r}\frac{V}{p+q}\hat{\Omega}_{\sigma\gamma\tau}(\bm{r})\times \nonumber \\
&&\left[p\frac{\phi_{\tau}(\bm{r})\phi_{\tau}(\bm{r})}{UN^2}+q\frac{\bar{\Delta}_{\tau}(\bm{r})\Delta_{\tau}(\bm{r})}{UN^2} \right]
\end{eqnarray}
where $U$ is the strength of trial interaction [\onlinecite{Macdonald,Kerman1,Kerman2,Kerman3}]. Here $U$ is set to be $U=V$. $N^2$ stands for a repeat sum over the $N$ single-particle state labels in the Hilbert space. $p$ and $q$ are two positive real numbers, and the results are independent of the choices of $p$ and $q$. For simplicity, we set $p=1$， $q=2$.

If we ignore the path integral over bosonic fields, the interaction term is in high order of $\epsilon$, which can be dropped, then we recover the mean field approximation in Sec II.  The mean field potential \{$\Delta_0$,$\bar{\Delta}_0$,$\phi_0$\} are given by the minimization of the mean field action $S_0$[\onlinecite{Macdonald}]. Going beyond the mean field theory, we restore the path integral and write the bosonic fields as $\Delta_{\tau}(\bm{r})=\Delta_0+\eta_{\tau}(\bm{r})$, $\bar{\Delta}_{\tau}(\bm{r})=\bar{\Delta}_0+\bar{\eta}_{\tau}(\bm{r})$, $\phi_{\tau}(\bm{r})=\phi_0+\xi_{\tau}(\bm{r})$. Expanding the action around its minimum up to the second order of $\eta$, $\bar{\eta}$ and $\xi$, we obtain

\begin{equation}
S\approx S_0+\frac{1}{2}\sum_{\tau,\tau'}\int d^2\bm{r}d^2\bm{r}'\bar{\Phi}_{\tau}(\bm{r})M_{\tau,\tau'}(\bm{r},\bm{r}')\Phi_{\tau'}(\bm{r}')
\end{equation}
where the bosonic fields are defined by $\bar{\Phi}=(\xi,\bar{\eta},\eta)^{\text{T}}$, $\Phi=(\xi,\eta,\bar{\eta})$. For convenience, we also define the bosonic operators $\hat{\Phi}=(\hat{\xi},\hat{\eta},\hat{\eta}^{\dag})^{\text{T}}$ with $\hat{\xi}(\bm{r})=\frac{1}{\sqrt{2}}\sum_{s\sigma\gamma}c_{s\sigma\gamma}^{\dag}(\bm{r})c_{s\sigma\gamma}(\bm{r})$ and $\hat{\eta}(\bm{r})=\sum_{\sigma\gamma}c_{\uparrow\sigma\gamma}(\bm{r})c_{\downarrow\bar{\sigma}\gamma}(\bm{r})$. The matrix $M$ is given by

\begin{eqnarray}
&&M_{\tau\tau'}^{ij}(\bm{r},\bm{r}')=\frac{\partial^2 S}{\partial\bar{\Phi}_{\tau}^i(\bm{r})\partial\Phi_{\tau'}^{j}(\bm{r}')} \nonumber \\
&&=\epsilon\left[\delta_{\tau,\tau'}U^{-1}+\epsilon(1-\delta_{\tau\tau'})D_{\tau\tau'}+\epsilon\delta_{\tau\tau'}S\right]
\end{eqnarray}
Here only time indices are explicitly shown. The matrices $D$ and $S$ are defined by

\begin{equation}
D_{\tau\tau'}^{ij}(\bm{r},\bm{r}')=\langle\text{T}_{\tau}\hat{\Phi}_{\tau}^{i}(\bm{r})\hat{\Phi}_{\tau'}^{j\dag}(\bm{r}')\rangle_{0}-\langle\hat{\Phi}_{\tau}^i(\bm{r})\rangle_{0}\langle\hat{\Phi}_{\tau'}^{j\dag}(\bm{r}')\rangle_{0}
\end{equation}

\begin{eqnarray}
S_{\tau}^{ij}(\bm{r},\bm{r}')&=&\frac{\delta_{ij}\delta(\bm{r}-\bm{r}') V\sum_{\sigma\gamma}\int d^2\bm{r}\langle\hat{\Omega}_{\sigma\gamma\tau}(\bm{r}) \rangle}{3UN^2}     \nonumber \\
&+&\langle\hat{\Phi}_{\tau}^i(\bm{r})\rangle_{0}\langle\hat{\Phi}_{\tau'}^{j\dag}(\bm{r}')\rangle_{0}
\end{eqnarray}
where $\langle\quad\rangle_0$ means mean field thermal dynamical average. As explained in Ref.[\onlinecite{Kerman2,Kerman3}], $S$ matrix represents the contribution of the single quasi-particle motion which remains beyond the mean field grand potential. Here we focus on the rest part which represents the fluctuation grand potential. Apparently, in our system $D$ is translational invariant. We can evaluate the matrix $D$ by doing Fourier transformation, which gives

\begin{eqnarray}\label{Eq:Dmatrix}
D_{11}(i\omega_m,\bm{q})&=&\frac{-2\Delta_0^2X}{E[4E^2-(i\omega_m)^2]}  \nonumber \\
D_{22}(i\omega_m,\bm{q})&=&D_{33}(-i\omega_m,-\bm{q}) \nonumber \\
&=&\frac{X(i\tilde{\mu}\omega_m-E^2-\tilde{\mu}^2)}{E[4E^2-(i\omega_m)^2]}  \nonumber \\
D^{12}(i\omega_m,\bm{q})&=&D^{13}(-i\omega_m,-\bm{q})\nonumber \\
&=&D^{21}(i\omega_m,\bm{q})\nonumber\\
&=&D^{31}(-i\omega_m,-\bm{q}) \nonumber \\
&=&\frac{X\Delta_0(-i\omega_m-2\tilde{\mu})}{\sqrt{2}E[4E^2-(i\omega_m)^2)]}  \nonumber \\
D_{23}(i\omega_m,\bm{q})&=&D_{32}(i\omega_m,\bm{q})\nonumber \\
&=&-\frac{X\Delta_0^2}{E(4E^2-(i\omega_m)^2)}
\end{eqnarray}
where $X=2N_{\phi}\tanh(E/2T)e^{-l_B^2\bm{q}^2/2}$ and $l_B$ is the pseudomagnetic length $l_B=\sqrt{\frac{1}{eB_p}}$. The calculation of $D$ matrix can be found in Appendix B.

Now, it is straightforward to work out the path integral. The fluctuation grand potential is given by

\begin{eqnarray}
\Omega_{fl}&=&\frac{T}{2}\text{Tr}\log(\mathbb{I}+UD)-\frac{1}{2}\text{Tr}(UD)\nonumber \\
&=&T\sum_{\omega_{\bm{q}}>0}\log\sinh(\omega_{\bm{q}}/2T)\nonumber\\&&-T\sum_{\bm{q}}\log\sinh(E/T)\nonumber\\
&&-\frac{1}{2}\text{Tr}(UD)
\end{eqnarray}
where $\omega_{\bm{q}}$ is collective excitation energy obtained by solving $\text{Det}(\mathbb{I}+UD)=0$. Combining Eq.(\ref{Eq:Dmatrix}) with the gap equation $\frac{1}{U}=\frac{N_{\phi}\tanh(E/2T)}{E}$, we can easily get $\omega_{\bm{q}}=2E(1-e^{-l_B^2\bm{q}^2/2})$. This system has only one well defined collective mode, which is Goldstone mode due to spontaneous symmetry breaking. At $\bm{q}=0$, this mode is purely due to pairing phase fluctuation. Away from $\bm{q}=0$ this mode is a mixture of phase, amplitude and density fluctuations. The second term comes from the scattering of gapped quasi-particles. The third term is to cancel the linear term in the expansion in powers of interaction strength [\onlinecite{Kerman2,Kerman3}]. The total fluctuation free energy is given by

\begin{equation}
\begin{aligned}
\Omega_{fl}=\sum_{\bm{q}}\left[T\log\sinh(\frac{E(1-e^{-l_B^2\bm{q}^2/2})}{T})\right.\\
\left.-T\log\sinh(E/T)+\frac{E}{\tanh(E/T)}e^{-l_B^2\bm{q}^2/2}\right]
\end{aligned}
\end{equation}

In order to calculate the critical temperature, we need to solve the gap equation and number equation simutaneously. Following the NSR method [\onlinecite{NSR,Randeria,Liu}], we keep the gap equation at mean field level while adding the fluctuation free energy into the number equation

\begin{eqnarray}\label{Eq:n}
n_e=-\frac{\partial\Omega}{\partial\mu}=-\frac{\partial\Omega_0}{\partial\mu}-\frac{\partial\Omega_{fl}}{\partial\mu}
\end{eqnarray} 
The first term in Eq.(\ref{Eq:n}) is the particle number from mean field grand potential $n_M=-\frac{\partial\Omega_0}{\partial\mu}=\frac{\mathcal{N}N_{\phi}[E+\tilde{\mu}\tanh(E/2T]}{E}$ where $\mathcal{N}$ is the number of unit cells. The second term is  particle number from fluctuations $n_{fl}=-\frac{\partial\Omega_{fl}}{\partial\mu}$, which can be calculated numerically.  Different from the tranditional NSR method [\onlinecite{NSR,Randeria,Liu}], we here take into account both density and pairing fluctuations. The advantage is that the density fluctuations can effectively describe the interaction between Cooper pair molecules, which is crucial in two dimensions. We then calculate the order parameters by solving the gap equation and number equation numerically and compare the critical temperatures as and order parameters with the results from the mean field approximation, shown in Fig.2. 

\begin{figure}[h!]
\centering
\subfigure{
\includegraphics[width=0.45\columnwidth, height=1.35in]{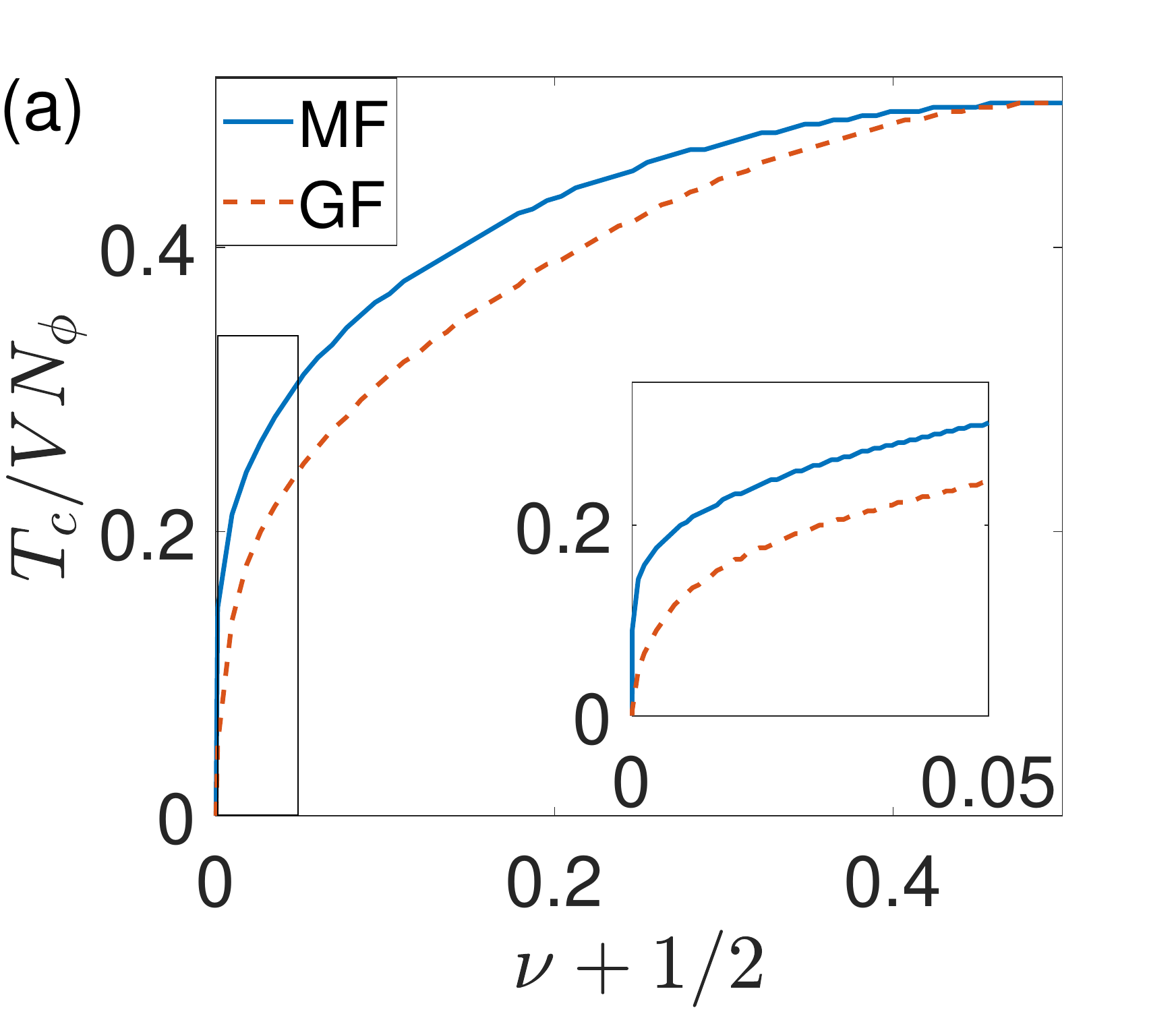}
}
\subfigure{
\includegraphics[width=0.45\columnwidth, height=1.35in]{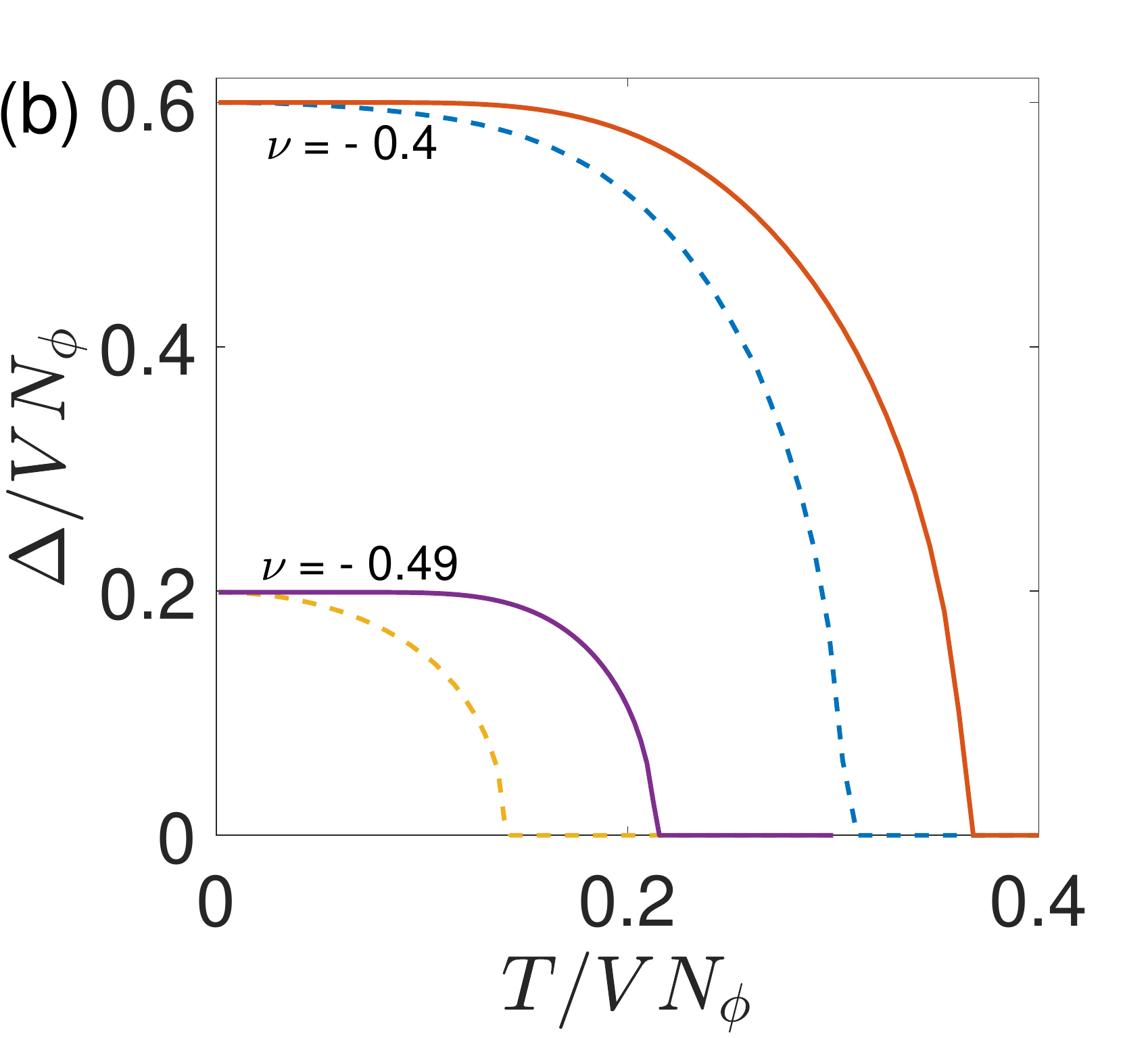}
}
\caption{(a) The critical temperature $T_c$ as a function of the filling factor $\nu$. The inserted figure is the zoom-in view of the rectangular region.  (b) The order parameter as a function of temperature for two different fillings as stated in the figure. In both figures, the dashed curves represent the results calculated with Gaussian fluctuations (GF) effect while the solid curves are calculated from merely mean field theory (MF).}\label{Fig:GF_MF}
\end{figure}

From Fig.\ref{Fig:GF_MF}(a), one can see that the Gaussian fluctuation effects reduce the critical temperatures. When the band is nearly empty, the ratio between mean field particle number and fluctuation particle number will approach zero at critical temperature, e.g. when $\nu = -0.499$, $n_M/n_{fl}=0.0065$ .  This means that in the dilute limit, the superconductivity is destroyed by collective motion of molecules at critical point. In this case, preformed Cooper pairs can still exist above critical temperatures. At exact half filling point, the fluctuations will not affect the critical temperature from our calculation. And from Fig.\ref{Fig:GF_MF}(b), it can be seen that the order parameters at zero temperature are the same for both cases --with and without fluctuation effects. This is consistent with the statement in Ref.[\onlinecite{Torma,Torma2}] that the mean field ground state is the exact ground state in a flat band superconductor. However at finite temperature,  the order parameter is hugely depressed by the fluctuation effects in the dilute limit(e.g. $\nu=-0.49$).

\section{Conclusion and Discussion}

In conclusion, we investigate the properties of a flat band superconductor realized with pseudomagnetic field preserving TRS. We propose this can be realized with molecular graphene on the top of a superconductor in the presence of three coplanar strain fields.  Applying mean-field approximation on the effective two dimensional Dirac Hamiltonian, we show that the pseudomagnetic fields enhance the orbital limit of the flat band superconductor. Here we only consider the orbital magnetization from external fields. When a $real$ magnetic field is applied perpendicularly to the sample, the superconducting phase can be killed by the Zeeman effect. Hence, the critical magnetic field of our model would be Pauli limit [\onlinecite{Clogson,Chandrasekhar}]. Fortunately, the Pauli limit in $z$ direction can be increased by Rashba spin orbital coupling in the superconducting substrate [\onlinecite{Sigrist}]. In this sense, a true high critical field superconductor can be achieved.

We further study fluctuation effects of flat band superconductors using the extended NSR method. Taking Gaussian fluctuations into account, we found the critical temperatures much lower than mean field results in the dilute limit. This can be understood that in the dilute limit the hardcore nature of Cooper pair molecules is not important and the system can be described by a weakly interacting boson model [\onlinecite{NSR}]. Thus the critical temperature is boson condensation temperature instead of pairing breaking temperature obtained from mean field calculation. When the filling factor is large, in contrast to the dilute case, the Gaussian fluctuations almost have no effect on the critical temperature. Notice that in our method, we treat  density fluctuations and pairing fluctuations on the equal footing, different from  Ref.[\onlinecite{NSR,Randeria,Liu}]. By doing this, we obtained zero fluctuation energy at zero temperature, in agreement with the previous theory [\onlinecite{Torma,Torma2}].

We comment also that long range Coulomb interaction, which may destroy superconductivity, is not considered in our model. One possible way to avoid this problem is to put the sample on substrates which screen Coulomb interaction at length scales larger than the pseudomagnetic length $l_B$. Lastly, we note that although here we start from a specific model, our analysis on fluctuation effects can be easily generalized to other flat band models. It may have possible application to the recently discovered flat band system in twisted bilayer graphene [\onlinecite{Cao1,Cao2}].

\section*{ACKNOWLEDGEMENTS} 

X. H. Li and Y. Lu thank the support of HKRGC through HKUST3/CRF/13G and C6026-16W.

\onecolumngrid
\appendix

\section{Derivation of Eq. (\ref{Eq:Gap2})}

We start with the BdG Hamiltonian.

\begin{equation}
h_{BdG}(\Delta=0)=\begin{pmatrix}
h_0   &&   0  \\
0  && -\Theta h_0\Theta^{-1}
\end{pmatrix}
\end{equation}
Note that $h_{BdG}$ commutes with the operator $P=s_z\tau_z\gamma_z$. Thus the Hamiltonian can be block diagonalized if we choose the basis to be the eigenvectors of $P$ operator.

\begin{equation}
\tilde{h}_{BdG}=\begin{pmatrix}
h_{BdG1}  &&  0  &&  0  &&  0  \\
0  &&  h_{BdG2}  &&  0  &&  0  \\
0  &&  0  &&  h_{BdG3}  &&  0  \\
0  &&  0  &&  0  &&  h_{BdG4}  \\ 
\end{pmatrix}
\end{equation}
where $h_{BdG1}$, $h_{BdG2}$, $h_{BdG3}$ and $h_{BdG4}$ are associated with each other by time reversal symmetry and particle hole symmetry. They will give the same results for orbital limit. Thus we can only consider $h_{BdG1}$. At the critical point, the order parameter vanishes, and $h_{BdG1}$ is given by

\begin{equation}
h_{BdG1}\equiv \begin{pmatrix}
h_e && 0 \\
0   && h_h
\end{pmatrix} 
=\begin{pmatrix}
\sigma_x(-i\nabla_x+eA_x)+\sigma_y(-i\nabla_y+eA_y) && 0 \\
0 && \sigma_x(-i\nabla_x-eA_x)+\sigma_y(-i\nabla_y-eA_y)
\end{pmatrix}
\end{equation}
Here we use $h_e$ and $h_h$ to denote electron Hamiltonian and hole Hamiltonian. We choose the Landau gauge for pseudomagnetic fields $A_{p,x}=yB_p,A_{p,y}=0$. In x direction, the Hamiltonian is periodical. Thus we can use LL index $n$ and electron momentum $k_x$ to label one single particle eigenstate of the Hamiltonian. The Green's function is given by
 
 \begin{equation}
  G_0(i\omega_m)\equiv \begin{pmatrix}
  G_{0e} && 0\\
  0 && G_{0h}
  \end{pmatrix}
  =\begin{pmatrix}
  \frac{1}{i\omega_m-h_e} && 0\\
  0 && \frac{1}{i\omega_m-h_h}
  \end{pmatrix}
 \end{equation}
The electron Green's function $G_{0e}$ and hole Green's function $G_{0h}$ can be written in terms of the eigenstates of $h_e$ and $h_h$. As the wave function of 0$th$ LL is localized at A sublattice, in the following we only write the Green's function for A sublattice

\begin{equation}
G_{0e}(i\omega_m)=\sum_{k_x,n}\frac{|n,k_x\rangle\langle n,k_x|}{i\omega_m-\epsilon_n}\approx \sum_{k_x}\frac{|0,k_x\rangle\langle 0,k_x|}{i\omega_m+\tilde{\mu}}
\end{equation}

\begin{equation}
G_{0h}(i\omega_m)=\sum_{k_x,n}\frac{|n,k_x\rangle\langle n,k_x|}{i\omega_m+\epsilon_n}\approx \sum_{k_x}\frac{|0,k_x\rangle\langle 0,k_x|}{i\omega_m-\tilde{\mu}}
\end{equation}
and

\begin{equation}\label{Eq:G0e0}
G_{0e}(i\omega_m,\bm{r}_1,\bm{r}_2)\approx \sum_{k_x}\frac{\langle \bm{r}_1|0,k_x\rangle\langle 0,k_x|\bm{r}_2\rangle}{i\omega_m+\tilde{\mu}}
\end{equation}

\begin{equation}
G_{0h}(i\omega_m,\bm{r}_1,\bm{r}_2)\approx \sum_{k_x}\frac{\langle \bm{r}_1|0,k_x\rangle\langle 0,k_x|\bm{r}_2\rangle}{i\omega_m-\tilde{\mu}}
\end{equation}
Here we use $|n,k\rangle$ to denote the state with LL index $n$ and momentum $k$. As we mentioned in section II, we focus on the 0$th$ LL and drop the higher ones. The 0$th$ LL wave function is given by

\begin{eqnarray}\label{Eq:WF}
\langle \bm{r}_1|0,k_x\rangle&=&\frac{1}{C}e^{-(r_{1,y}/l_B-k_xl_B)^2/2}e^{ik_xr_{1,x}} \nonumber \\
\langle 0,k_x|\bm{r}_2\rangle&=&\frac{1}{C}e^{-(r_{2,y}/l_B-k_xl_B)^2/2}e^{-ik_xr_{2,x}} 
\end{eqnarray}
where $l_B$ is the pseudomagnetic length defined by $l_B=\sqrt{\frac{1}{B_p}}$ and $C$ is the constant normlization factor. Substituting Eq.(\ref{Eq:WF}) into Eq. (\ref{Eq:G0e0}), we have

\begin{equation}
G_{0e}(i\omega_m,\bm{r}_1,\bm{r}_2)=\frac{1}{(i\omega_m+\tilde{\mu})C^2} \sum_{k_x}\exp[-(r_{1,y}/l_B-k_xl_B)^2/2-(r_{2,y}/l_B-k_xl_B)^2/2+ik_x(r_{1,x}-r_{2,x})]
\end{equation}
Replacing $k_x$ by $k_x-(\bm{r}_{1,y}+\bm{r}_{2,y})/2l_B^2$, $G_{0e}$ can be written as

\begin{eqnarray}
G_{0e}(i\omega_m,\bm{r}_1,\bm{r}_2)&=&\frac{1}{(i\omega_m+\tilde{\mu})C^2}\sum_{k_x}\exp[-((r_{1,y}-r_{2,y})/2l_B-k_xl_B)^2/2 \nonumber \\
&&-(-(r_{1,y}-r_{2,y})/2l_B-k_xl_B)^2/2+i(k_x-(r_{1,y}+r_{2,y})/2l_B^2)(r_{1,x}-r_{2,x})]
\end{eqnarray}
Defining $r_{+/-,x/y}=r_{1,x/y}\pm r_{2,x/y}$ and working out the sum over $k_x$, we arrive at

\begin{eqnarray}\label{Eq:G0e}
G_{0e}(i\omega_m,\bm{r}_1,\bm{r}_2)=\frac{1}{(i\omega_m+\tilde{\mu})\tilde{C}}\exp[-(r_{-,x}^2+r_{-,y}^2)/4l_B^2-ir_{+,y}r_{-,x}/2l_B^2)]
\end{eqnarray}

In the same method, we get $G_{0h}$

\begin{eqnarray}\label{Eq:G0h}
G_{0h}(i\omega_m,\bm{r}_1,\bm{r}_2)=\frac{1}{(i\omega_m-\tilde{\mu})\tilde{C}}\exp[-(r_{-,x}^2+r_{-,y}^2)/4l_B^2-ir_{+,y}r_{-,x}/2l_B^2)]
\end{eqnarray}
where $\tilde{C}$ is just a constant. In terms of $G_{0e}$ and $G_{0h}$, Eq (\ref{Eq:SemiGap}) can be written as

\begin{equation}\label{Eq:GapG}
\Delta(\bm{r}_2)=-2TV\sum_m\int d^2\bm{r}_1G_{0e}(i\omega_m,\bm{r}_1,\bm{r}_2)G_{0h}(-i\omega_m,\bm{r}_2,\bm{r}_1)e^{(\bm{r}_2-\bm{r}_1)(\bm{\nabla}-2ie\bm{A}_{r})}\Delta(\bm{r}_2)
\end{equation}
Substituting Eq. (\ref{Eq:G0e}) and (\ref{Eq:G0h}) into Eq. (\ref{Eq:GapG}), we have

\begin{equation}\label{Eq:Gap3}
\Delta(\bm{r})=-TV\sum_m\int d\rho \frac{2\rho e^{-\rho^2/2l_B^2}e^{\rho(\bm{\nabla}-2ie\bm{A}_{r})}\Delta(\bm{r})}{(i\omega_m+\tilde{\mu})(i\omega_m-\tilde{\mu})\tilde{C}^2}
\end{equation}
where $\rho$ is defined by $\rho=|\bm{r}_1-\bm{r}_2|$. Eq.(\ref{Eq:Gap3}) can be further  simplified following [\onlinecite{Huxley}]. Working out the frequency summation, we have

\begin{eqnarray}\label{Eq:Gap4}
\Delta(\bm{r})=V\int d\rho  Y\rho\tanh(\tilde{\mu}/2T)e^{-\frac{1}{2}\rho^2eB_p}e^{-\frac{1}{2}\rho^2eB_c}\Delta(\bm{r})
\end{eqnarray}
The constant prefactor $Y$ can be obtained by solving Eq. (\ref{Eq:Gap4}) at $B_c=0$,  $\frac{1}{V}=\frac{N_{\phi}\tanh(\tilde{\mu}/2T)}{\tilde{\mu}}$. Finally we have

\begin{eqnarray}
\Delta(\bm{r})=V\int d\rho \frac{\rho N_{\phi}}{\tilde{\mu}}\tanh(\tilde{\mu}/2T)e^{-\frac{1}{2}\rho^2eB_p}e^{-\frac{1}{2}\rho^2eB_c}\Delta(\bm{r})
\end{eqnarray}
which is Eq. (\ref{Eq:Gap2}). 

\section{Calculation of $D$ Matrix}

Here we first calculate $D^{11}$ for example. As mentioned in Appendix. A, we can block diagonalize the Hamiltonian by rearranging the basis. This implies that we can consider only $c_{\uparrow,K,A}$ and $c_{\downarrow,K',A}$ electrons. Then we multiply the result by 4 accounting for the valley degeneracy. For simplicity, in this appendix we use $c_{\uparrow}\equiv c_{\uparrow,K,A}$ and $c_{\downarrow}\equiv c_{\downarrow,K',A}$. Thus we have

\begin{eqnarray}
D_{\tau\tau'}^{11}(\bm{r},\bm{r}')&=&\langle\text{T}_{\tau}\hat{\xi}_{\tau}(\bm{r})\hat{\xi}_{\tau'}(\bm{r}')\rangle_0-\langle\hat{\xi}_{\tau}(\bm{r})\rangle_0\langle\hat{\xi}_{\tau'}(\bm{r}')\rangle_0 \nonumber \\
&=&4\langle\text{T}_{\tau} c_{\uparrow\tau}^{\dag}(\bm{r})c_{\uparrow\tau}(\bm{r})c_{\uparrow\tau'}^{\dag}(\bm{r}')c_{\uparrow\tau'}(\bm{r}')\rangle_0-4\langle c_{\uparrow\tau}^{\dag}(\bm{r})c_{\uparrow\tau}(\bm{r})\rangle_0\langle c_{\uparrow\tau'}^{\dag}(\bm{r}')c_{\uparrow\tau'}(\bm{r}')\rangle_0 \nonumber \\
&=&4\langle\text{T}_{\tau}c_{\uparrow,\tau}(\bm{r})c_{\uparrow,\tau'}^{\dag}(\bm{r}')\rangle_0\langle\text{T}_{\tau}c_{\uparrow,\tau'}(\bm{r}')c_{\uparrow,\tau}^{\dag}(\bm{r})\rangle_0 
\end{eqnarray}
Let $G_{\tau\tau'}^{ee}(\bm{r},\bm{r}')\equiv-\langle\text{T}_{\tau}c_{\uparrow\tau}(\bm{r})c_{\uparrow\tau'}^{\dag}(\bm{r}')\rangle_0$ be the normal Green's function. Doing Fourier transformation, we find the Green's function in momentum and frequency space

\begin{eqnarray}
G^{ee}(i\omega_n,\bm{k})=\int_0^{\beta} d\tau\int d^2rG_{\tau\tau'}^{ee}(\bm{r},\bm{r}')e^{i\omega_n\tau-i\bm{k}\bm{r}}=\frac{-i\omega_n+\tilde{\mu}}{E^2-(i\omega_n)^2}
\end{eqnarray}
The Fourier transformation of $D^{11}$ gives

\begin{eqnarray}
D^{11}(i\omega_m,\bm{q})&=&\int_0^{\beta}d\tau\int d^2r D_{\tau\tau'}^{ee}(\bm{r},\bm{r}')e^{i\omega_m\tau-i\bm{q}\bm{r}}  \nonumber \\
&=& 4\sum_{n,\bm{k}} G^{ee}(i\omega_n+i\omega_m,\bm{k}+\bm{q})G^{ee}(i\omega_n,\bm{k})e^{-l_B^2\bm{q}^2/2} \nonumber \\
&=&4\sum_{n,\bm{k}}\frac{(-i\omega_n-i\omega_m+\tilde{\mu})}{[E^2-(i\omega_m+i\omega_n)^2]}\frac{(-i\omega_n+\tilde{\mu})}{[E^2-(i\omega_n)^2]}e^{-l_B^2\bm{q}^2/2} \nonumber \\
&=&-\frac{2X\Delta_0^2}{E[4E^2-(i\omega_m)^2]}
\end{eqnarray}
where $X=2N_{\phi}\tanh(E/2T)e^{-l_B^2\bm{q}^2/2}$. To calculate other elements in $D$ matrix, we define the following Green's functions 

\begin{eqnarray}
G_{\tau\tau'}^{he}(\bm{r},\bm{r}')&=&-\langle \text{T}_{\tau}c_{\downarrow,\tau}^{\dag}(\bm{r})c_{\uparrow,\tau'}^{\dag}(\bm{r}')\rangle_0 \nonumber \\
G_{\tau\tau'}^{eh}(\bm{r},\bm{r}')&=&-\langle \text{T}_{\tau}c_{\uparrow,\tau}(\bm{r})c_{\downarrow,\tau'}(\bm{r}')\rangle_0 \nonumber\\
G_{\tau\tau'}^{hh}(\bm{r},\bm{r}')&=&-\langle\text{T}_{\tau}c_{\uparrow,\tau}^{\dag}(\bm{r})c_{\downarrow,\tau'}(\bm{r}')\rangle_0 
\end{eqnarray}
and the Fourier transformation

\begin{eqnarray}
G^{eh}(i\omega_n,\bm{k})&=&\int_0^{\beta} d\tau\int d^2\bm{r}G_{\tau\tau'}^{eh}(\bm{r},\bm{r}')e^{i\omega_n\tau-i\bm{k}\bm{r}}=\frac{-\Delta_0}{E^2-(i\omega_n)^2} \nonumber \\
G^{he}(i\omega_n,\bm{k})&=&\int_0^{\beta} d\tau\int d^2\bm{r}G_{\tau\tau'}^{he}(\bm{r},\bm{r}')e^{i\omega_n\tau-i\bm{k}\bm{r}}=\frac{-\Delta_0}{E^2-(i\omega_n)^2} \nonumber \\
G^{hh}(i\omega_n,\bm{k})&=&\int_0^{\beta} d\tau\int d^2\bm{r}G_{\tau\tau'}^{hh}(\bm{r},\bm{r}')e^{i\omega_n\tau-i\bm{k}\bm{r}}=\frac{-i\omega_n-\tilde{\mu}}{E^2-(i\omega_n)^2} \nonumber \\
\end{eqnarray}
 
In the same method as we calculate $D^{11}$, we can obtain all the elements in $D$ matrix

\begin{eqnarray}
D^{22}(i\omega_m,\bm{q})&=&D^{33}(-i\omega_m,-\bm{q})\nonumber \\
&=& 2\sum_{n,\bm{k}} G^{ee}(i\omega_n+i\omega_m,\bm{k}+\bm{q})G^{hh}(i\omega_n,\bm{k})e^{-l_B^2\bm{q}^2/2} \nonumber \\
&=&2\sum_{n,\bm{k}}\frac{(-i\omega_n-i\omega_m+\tilde{\mu})}{[E^2-(i\omega_m+i\omega_n)^2]}\frac{(-i\omega_n-\tilde{\mu})}{[E^2-(i\omega_n)^2]}e^{-l_B^2\bm{q}^2/2} \nonumber \\
&=&\frac{X(i\tilde{\mu}\omega_m-E^2-\tilde{\mu}^2)}{E(4E^2-(i\omega_m)^2)} 
\end{eqnarray}

\begin{eqnarray}
D^{12}(i\omega_m,\bm{q})&=&D^{13}(-i\omega_m,-\bm{q})\nonumber \\
&=&D^{21}(i\omega_m,\bm{q})\nonumber\\
&=&D^{31}(-i\omega_m,-\bm{q}) \nonumber \\
&=& 2\sqrt{2}\sum_{n,\bm{k}} G^{ee}(i\omega_n+i\omega_m,\bm{k}+\bm{q})G^{eh}(i\omega_n,\bm{k})e^{-l_B^2\bm{q}^2/2} \nonumber \\
&=&2\sqrt{2}\sum_{n,\bm{k}}\frac{(-i\omega_n-i\omega_m+\tilde{\mu})}{[E^2-(i\omega_m+i\omega_n)^2]}\frac{(-\Delta_0)}{[E^2-(i\omega_n)^2]}e^{-l_B^2\bm{q}^2/2} \nonumber \\
&=&\frac{X\Delta_0(-i\omega_m-2\tilde{\mu})}{\sqrt{2}E[4E^2-(i\omega_m)^2)]} 
\end{eqnarray}

\begin{eqnarray}
D^{32}(i\omega_m,\bm{q})&=&D^{23}(i\omega_m,\bm{q})\nonumber\\
&=&2\sum_{n,\bm{k}} G^{eh}(i\omega_n+i\omega_m,\bm{k}+\bm{q})G^{he}(i\omega_n,\bm{k})e^{-l_B^2\bm{q}^2/2} \nonumber \\
&=&2\sum_{n,\bm{k}}\frac{(-\Delta_0)}{[E^2-(i\omega_n+i\omega_m)^2]}\frac{(-\Delta_0)}{[E^2-(i\omega_n)^2]}e^{-l_B^2\bm{q}^2/2} \nonumber \\
&=&-\sum_{n,\bm{k}}\frac{\Delta_0^2}{E}\frac{e^{-l_B^2\bm{q}^2/2}}{[4E^2-(i\omega_m)^2]}\nonumber \\
&=&-\frac{X\Delta_0^2}{E[4E^2-(i\omega_m]^2)}
\end{eqnarray}

\end{document}